\documentclass[aps,pra,twocolumn]{revtex4}

\usepackage{graphicx}

\begin{document}
\title{\begin{flushright}\begin{tiny}UWthPh--2004--44\end{tiny}\end{flushright}
\vspace{-0.7cm} Thermodynamical Versus Optical Complementarity}

\author{Beatrix C. Hiesmayr$^1$ and Vlatko Vedral$^{2,3,4}$}
\address{$^1$Institut f\"ur Theoretische Physik, Boltzmanngasse 5, 1090 Vienna, Austria}
\address{$^2$Institut f\"ur Experimental Physik, Boltzmanngasse 5, 1090 Vienna,
Austria}
\address{$^3$Erwin Schr\"odinger Institut, Boltzmanngasse 9, 1090 Vienna,
Austria}
\address{$^4$The School of Physics and Astronomy, University of Leeds, LS2 9JT,
United Kingdom}
\date{\today}
\begin{abstract}
We establish correspondence between macroscopic thermodynamical quantities and
complementarity in wave interference. The well known visibility and
predictability in a double slit--like experiment are shown to be connected to
magnetic susceptibility and magnetization of a general interacting spin chain.
This gives us the ability to analyze the tradeoff between thermodynamical
quantities in the same information--theoretic way that is used in analyzing the
wave--particle duality. We thus obtain new physical insights into usually
complicated thermodynamical models, such as viewing a phase transition simply
as a change from an effective single slit diffraction to a double slit
interference. PACS-Numbers: 42.25Hz, 05.70.-a, 05.30.–d
\end{abstract}
\draft \maketitle

\noindent \textbf{Introduction.} Spin chains have been analyzed
within many contexts in the last hundred years of physics. They
frequently provide very precise models for solids within which one
can study their various macroscopic properties. Typically one
studies the response of solids to different external conditions,
such as the changing magnetic field, temperature etc. The standard
approach is first to construct the Hamiltonian for the chain, which
is then diagonalized through a sequence of transformations. The
resulting eigenvalues are then used to obtain the partition function
and from that all macroscopic quantities can in principle be
derived.

It frequently happens that the diagonalizing method involves highly complicated
transformations which do not reveal much physics, and ultimately leave very
little trace in the final quantities such as the heat capacity or magnetic
susceptibility. Physicists are not really interested in the details of
diagonalization, but in the relationship between various macroscopic
quantities, any tradeoff between them and any limitation in the information we
can have about them. Diagonalization procedure itself, however, throws very
little light on these issues. Partition functions, in particular, do not by
themselves reveal much about the relationships between derived macroscopic
quantities.

The main purpose of this Letter is to further our understanding of
thermodynamical quantities by making an analogy between optical interference
phenomena and the summation of the exponential Boltzmann energy factors in the
partition function (of spin lattice). As we will show this analogy helps us to
clarify relationships and tradeoffs between thermodynamical quantities such as
magnetization and magnetic susceptibility in the same way that we understand
the tradeoff between the interference fringes and the ``which--path"
information in an interferometer. Our work is also inspired by the analogy
between Feynman's quantum mechanical propagator and the thermodynamical
partition function \cite{Feynman}.

The Letter is organized as follows. We first describe complementarity in a
double slit experiment between the ``predictability" of the wave/particle going
through one of the slits and the ``visibility" of the interference fringes.
Predictability and visibility, which will be formally quantified below, will be
our pair of complementary properties, so that the better we know one of them,
the less we can determine the other one. We will then discuss the
correspondence between double slit interference and the thermodynamical
partition function for a chain of non-interacting two level systems. From that
we will argue that there is a thermodynamical analogue to the dynamical double
slit complementarity. This thermodynamical complementarity is seen in a
tradeoff relationship between quantities such as the magnetization of a system
and its susceptibility. Roughly speaking, the higher the value of one of them,
the lower will be the value of the other one, in direct analogy with
predictability and visibility. We show that this is a general result that
applies to more complicated interacting spin chains and allows us to understand
phase transitions in the same spirit.

\smallskip

\noindent \textbf{Complementarity in double slit experiments.} The
concept of duality in interferometric or double slit--like devices
is the basic ingredient of any physical theory of waves. It is also
at the heart of the quantum mechanical wave--particle duality, since
in quantum theory waves are used to describe matter as well. For
simplicity, we will use the quantum language to discuss
interference, however, the whole analysis applies to any (classical
or quantum) wave theory. The qualitative statement that
``\textit{the observation of an interference pattern and the
acquisition of which--way information are mutually exclusive}'' has
only recently been rephrased as a quantitative statement
\cite{Englert}:
\begin{equation}\label{comp}
{\cal P}^2(y)+{\cal V}_0^2(y)\leq 1\;.
\end{equation}
where the equal sign is only valid for pure states. ${\cal V}_0$ is
the fringe visibility which quantifies the sharpness or contrast of
the interference pattern (``the wave--like property''), whereas
${\cal P}$ denotes the the path predictability, i.e., the \textit{a
priori} knowledge one can have on the path taken by the interfering
system (``the particle--like property''). Since we restrict our
analysis to two-path interferometry, the predictability is defined
by
\begin{equation}
{\cal P}\;=\;|p_I-p_{II}|\;,
\end{equation}
where $p_I$ and $p_{II}$ are the probabilities for taking each path
($p_I+p_{II}=1)$. Usually these quantities dependent on one external
parameter  which we label by $y$. For example consider the double
slit experiment for which the intensity is given by
\begin{equation}
I(y)\;=\; F(y)\;\big(1+{\cal V}_0(y) \cos(\phi(y)\big) ,
\end{equation}
where $F(y)$ is specific for each setup and $\phi(y)$ is the phase-difference
between the two paths. The variable $y$ characterizes in this case the detector
position. Note that this formalism applies to many different physical
situations. In Ref.\cite{Beatrix} the authors investigated physical situations
for which the expressions of ${\cal V}_0(y), {\cal P}(y)$ and $\phi(y)$ can be
analytically computed, i.e. they depend only linearly on the variable $y$. This
included interference patterns of various types of double slit experiments ($y$
is linked to position), but also oscillations due to particle mixing ($y$ is
linked to time), e.g. by the neutral kaon system, and also Mott scattering
experiments of identical particles or nuclei ($y$ is linked to a scattering
angle). All these two-state systems belonging to distinct fields of physics can
then be treated via the generalized complementarity relation in a unified way.
\smallskip

\noindent \textbf{Thermodynamical complementarity.} We first
illustrate with a simple example how complementary manifests itself
in thermodynamical systems. Suppose that we have an ensemble of $N$
two-level non-interacting systems, with eigenvalues $E_1= E$ and
$E_2=-E$. Then the partition function is
\begin{equation}
Z = (e^{-E/kT} + e^{+E/kT})^N = 2^N \cosh^N \frac{E}{kT}\;.
\end{equation}
This contains all thermodynamical information one can extract from
the system. Let us first write down the free energy as
\begin{equation}
F = -kT \ln Z = -kT N \ln (2\cosh \frac{E}{kT})\;.
\end{equation}
Suppose that this describes $N$ spins in an external magnetic field
in which case the energy is proportional to the magnetic field $B$,
like so $E=\mu B$. Now, the magnetization and the susceptibility are
given by:
\begin{eqnarray}
M & = & \frac{\partial F}{\partial B} =  2\mu N \tanh
\frac{E}{kT}\label{mag}\\
\chi & = & \frac{\partial^2 F}{\partial B^2} =  2\frac{\mu^2 N}{kT}
\cosh^{-2} \frac{E}{kT}\;. \label{sus}
\end{eqnarray}
The crux of our analogy is that $M$ and $\chi$ behave like the
predictability and the visibility. More precisely, one immediately
derives from Eq.(\ref{mag}) and Eq.(\ref{sus}) that the following
equation has to hold
\begin{equation}
(\frac{M}{2 \mu N})^2 + 2kT \frac{\chi}{4 \mu^2 N} = 1\;.
\end{equation}
There is a tradeoff between magnetization and susceptibility (per
spin) at a fixed temperature: the larger one of the quantities, the
smaller the other one and vice versa. Physically, this is to be
expected, since higher magnetization implies a better knowledge of
the spin direction and, in turn, that different spins are less
correlated (in the statistical sense); hence susceptibility is
smaller.

In order to compare this with the wave complementarity (both
classical and quantum), we now discuss interference pattern from two
slits, whose amplitude transmission function is a Gaussian. The
intensity is given by
\begin{eqnarray*}
I(y) & \propto & \left|\,e^{-(y-d/2)^2/2\sigma^2} + e^{-(y+d/2)^2/2\sigma^2}e^{i\phi (y)} \right|^2\nonumber\\
& = & e^{-(y^2+d^2/4)/\sigma^2}2\cosh(y d/\sigma^2) \biggl\{1 +
\frac{\cos \phi (y)}{\cosh (y d/\sigma^2)} \biggr\}
\end{eqnarray*}
where $d$ is the separation of the slits, $\sigma$ is the effective
width of the amplitude transmission function and $\phi(y)$ is the
phase arising from the path difference. From this formula we can
infer that the visibility is given by
\begin{equation}
{\cal V}_0(y) = \frac{1}{\cosh (y d/\sigma^2)}\;.
\end{equation}
The predictability can likewise be derived:
\begin{equation}
{\cal P}(y) = |\tanh (y d/\sigma^2)|\; ,
\end{equation}
which obviously fulfills the complementarity relation (\ref{comp}) for all $y$
with the equality sign. We can see that the predictability has exactly the same
functional behavior as the magnetization (per spin) in Eq. (\ref{mag}). The
susceptibility in Eq. (\ref{sus}), on the other hand, behaves in the same way
as the visibility. Pushing this analogy one step further, we can identify the
energy of the spins with the position variable in the double slit experiment
and likewise the inverse of the Boltzmann constant $k$ with the slit separation
$d$. In this case, the temperature $T$ corresponds to the effective slit width
$\sigma^2$ and the energy $E$ would be the detector position $y$. For fixed
energy, if the temperature is low (corresponding to a narrow slit), the
visibility goes to zero, and so (from the existing complementarity) the
predictability achieves the maximum value of one. We can also make a different
correspondence altogether that will be useful in our interferometric
understanding of the phase transition. The ratio of spin energy to temperature
$E/T$ could correspond to the detector position $y$, while $1/k$ is the ratio
$d/\sigma^2$.
\begin{figure}
\center{\includegraphics[width=8.5cm,keepaspectratio=true]{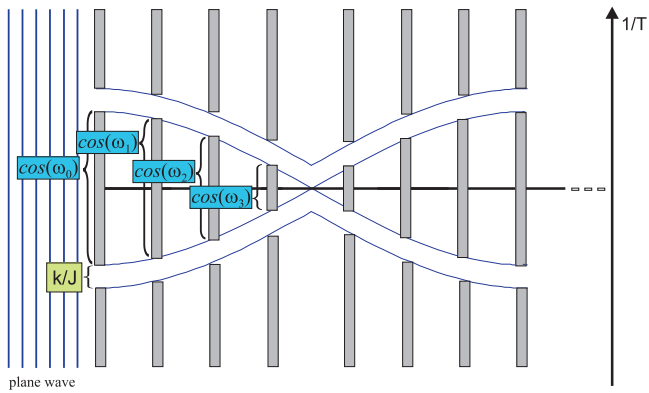}}
\caption{This picture shows an array of double slits which in our
correspondence is the optical analog to the partition function of
the $XY$ model. The separation of the two slits varies with the
$\cos$ of the different frequencies $\omega_k=2\pi k/N$ in the
partition function and the effective slit width is given by the
Boltzmann constant $k$ divided by the coupling $J$. The
thermodynamical limit is obtained by making the array continuous in
which case the only surviving interference is between the upper and
lower path (all other cross terms cancel). This then reproduces the
integral in Eq.(\ref{XYpartition}). In one analogy  different
detector positions at the screen correspond to different reciprocal
temperatures. Therefore if the temperature is high, we have the
central maximum of the interference pattern where both of the paths
contribute equally (thermodynamically this means that both Boltzmann
factors are equal). Remarkably, interference of Boltzmann factors
may exist even at zero temperature, and this then signifies the
point of quantum phase transition as explained in more detail in the
text.}\label{fig1}
\end{figure}

%%%%%%%%%%%%%%%%%%%%%%%%%%%%%%%%%%%%%%%%%%%%%%%%%%%%%%%%%%%%%%%%%%%%%%%%%%%%%%%%%%%%%%%%%%%%%%%%%%%%%%%%%%%%%%%%%%%%%%%%%
\smallskip

\noindent \textbf{General Relationship.}
%%%%%%%%%%%%%%%%%%%%%%%%%%%%%%%%%%%%%%%%%%%%%%%%%%%%%%%%%%%%%%%%%%%%%%%%%%%%%%%%%%%%%%%%%%%%%%%%%%%%%%%%%%%%%%%%%%%%%%%%%
We can formalize the tradeoff between the thermodynamical quantities
such as the (normalized) magnetization and susceptibility as well as
the optical quantities such as the predictability and visibility. A
general way of phrasing the complementarity between these is to say
that we have a function $f$ bounded between $-1$ and $1$ and we
compare its square with its derivative, $f'$. More precisely, we
would like to know for what functions $f$ do we have that
\begin{equation}
f^{2}(x/\alpha) + \alpha f'(x/\alpha) = \beta
\end{equation}
where for our physical applications $\alpha$ and $\beta$ are non-negative
constants. The most general solution is $f(x/\alpha)  = \sqrt{\beta}\tanh
\{\sqrt{\beta} (x/\alpha + c)\}$, where $c$ is another constant. We can see
that the most general solution of the above equation has the form of
predictability of Gaussians, or, alternatively, it has the form of a derivative
of partition function for a two level system. In the optical case, the
predictability is just the difference between the mod squares of the two
amplitudes for each slit. Since each amplitude is a Gaussian, this quantity
behaves like the tanh function. In thermodynamics, likewise, the partition
function for a two level system is a sum of the exponential Boltzmann factors
for the two states which is proportional to cosh. When this is differentiated
to obtain the magnetization, this gives us again the $\tanh$ function.

Our general complementarity bound is in fact an inequality of the type
$f^{2}(x/\alpha) + \alpha f'(x/\alpha) \le 1$. Some solutions of this
inequality can be constructed from the above equality, but we do not have the
most general closed form solution. Examples of these more general forms are now
shown to exist in well known lattice models.

%%%%%%%%%%%%%%%%%%%%%%%%%%%%%%%%%%%%%%%%%%%%%%%%%%%%%%%%%%%%%%%%%%%%%%%%%%%%%%%%%%%%%%%%%%%%%%%%%%%%%%%%%%%%%%%%%%%%%%%%%%
\smallskip

\noindent \textbf{The $XY$ Heisenberg Model.}
%%%%%%%%%%%%%%%%%%%%%%%%%%%%%%%%%%%%%%%%%%%%%%%%%%%%%%%%%%%%%%%%%%%%%%%%%%%%%%%%%%%%%%%%%%%%%%%%%%%%%%%%%%%%%%%%%%%%%%%%%%
This model has been extensively used in many situations and has been
analyzed in various regimes \cite{sach99}. The Hamiltonian is given
by:
\begin{eqnarray*}
H = -\frac{J}{2} \sum_{i=1}^{N} \sigma_i^x\otimes \sigma_{i+1}^x
+\sigma_i^y\otimes \sigma_{i+1}^y - \mu B \sum_{i=1}^{N} \sigma_i^z
\; .
\end{eqnarray*}
This Hamiltonian was diagonalized in \cite{Kat} by a sequence of
complicated transformations. In the large $N$ limit (which is what
we are always interested in when it comes to computing
thermodynamical quantities) one derives the following partition
function
\begin{equation}\label{XYpartition}
\lim_{N\longrightarrow \infty} \frac{1}{N}\ln(\frac{Z}{2^N})\; =\;
\frac{1}{\pi}
 \int_0^{\pi} \ln \cosh(C-2 K \cos(\omega)) d\omega
\end{equation}
where $K=J/2kT$, $C=\mu B/kT$ and $\omega=2\pi k/N$ are the frequencies of the
Fourier transformed fermionic creation and annihilation operators. The free
energy is
\begin{eqnarray*}
-\frac{F}{N k T}\;=\;\frac{1}{\pi}
 \int_0^{\pi} \ln 2 \cosh(C-2 K \cos(\omega)) d\omega\;,
\end{eqnarray*}
and so the magetization and the susceptibility are:
\begin{eqnarray*}
\frac{M}{N \mu}&=&\frac{1}{\pi}
 \int_0^{\pi} \tanh(C-2 K \cos(\omega)) d\omega\;, \\
\frac{\chi k T}{\mu^2 N}& =& \frac{1}{\pi}
 \int_0^{\pi} \cosh^{-2}(C-2 K \cos(\omega)) d\omega\;.
\end{eqnarray*}
Consequently, we obtain the following complementary relation
($\tilde{C}(\omega)=C-2 K \cos(\omega)$)
\begin{eqnarray*}
&&\biggr(\frac{M}{N \mu}\biggl)^2+\frac{\chi k T}{\mu^2
N}\;\leq\nonumber\\
&& \frac{1}{\pi}
 \int_0^{\pi}\tanh^2(\tilde{C}(\omega))+\cosh^{-2}(\tilde{C}(\omega))
 d\omega=1\;.
\end{eqnarray*}
We see that the form of these quantities is now more complicated. For each
$\omega$ the visibility is given by a $\cosh$-function corresponding to a
double slit experiment, however, the susceptibility per spin, i.e. the
measurable quantity, is the sum of all visibilities squared. This we can
illustrate by a series of double slits as in Fig.\ref{fig1} where we set for
simplicity $B=0$ and interpret the inverse temperature $1/T$ as the detector
position and $\frac{k}{J}$ as the constant effective slit width. Each of the
double slits in the series gives rise to its own predictability and the total
predictability is an integral over all individual ones. The integration is over
different slit separations corresponding to the value of $\cos(\omega)$.

%%%%%%%%%%%%%%%%%%%%%%%%%%%%%%%%%%%%%%%%%%%%%%%%%%%%%%%%%%%%%%%%%%%%%%%%%%%%%%%%%%%%%%%%%%%%%%%%%%%%%%%%%%%%%%%%%%%%%%%%%%
\smallskip

\noindent \textbf{Phase transitions.}
%%%%%%%%%%%%%%%%%%%%%%%%%%%%%%%%%%%%%%%%%%%%%%%%%%%%%%%%%%%%%%%%%%%%%%%%%%%%%%%%%%%%%%%%%%%%%%%%%%%%%%%%%%%%%%%%%%%%%%%%%%
Our correspondence between wave optics and thermodynamics allows us to
understand and interpret phase transitions in an elegant and transparent way.
Briefly stated, for a phase transition to occur in thermodynamics we need the
interference in the corresponding optical setting to appear from a pattern that
is otherwise effectively a single slit diffraction. If we never have
``interference'' between two different Boltzmann contributions then we can
conclude there is no phase transition for that thermodynamical model. The
simplest example where there is no interference is the one dimensional Ising
model whose Hamiltonian is given by
\begin{equation}
H=-J \sum_i \sigma_i ^z\sigma_{i+1}^z-\mu B\sum_i\sigma_i^z\;.
\end{equation}
The partition function has contributions of two
eigenvalues~$\lambda_+\ge \lambda_-$, however, in the large $N$
limit, only the larger of the two eigenvalues survives:
\begin{eqnarray}
F=\frac{1}{N} \log(\lambda_+^N+\lambda_-^N)
\stackrel{N\rightarrow\infty}\longrightarrow\log\lambda_+
\end{eqnarray}
This model is therefore never a two level system  and hence there is
no phase transition.
\begin{figure}
\center{
\includegraphics[width=8.5cm,keepaspectratio=true]{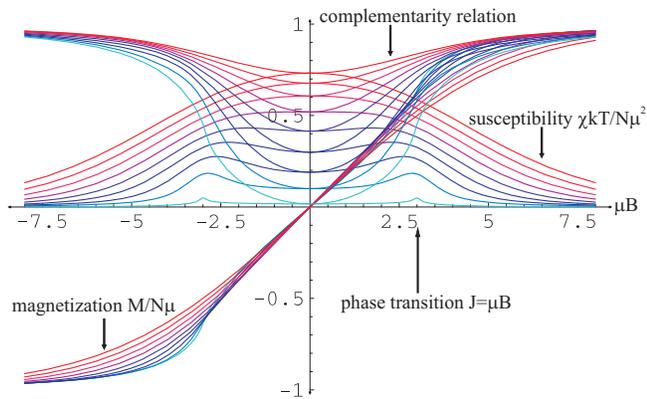}}
\caption{This figure shows the magnetization, susceptability and the
resulting complementarity relation of the transverse Ising model
depending on the magnetic field for different temperatures (blue
$\equiv k T$ small; coupling fixed to $J=3$). The quantum phase
transition occurs at $J=B$ for low temperature, where the
complementarity relation is most interesting because both terms
contribute. Away from this point either or both of the complimentary
quantities become small.}\label{fig2}
\end{figure}

On the other hand, the Ising model where the external field is in
the $x$ instead of the $z$ direction (a transverse field) behaves
very differently. Its free energy is given by:
\begin{eqnarray*}
\frac{-F}{N k T} = \frac{1}{\pi} \int_0^\pi \log( 2
\cosh\sqrt{K^2+C^2-2 K C \cos\omega}) d\omega\; ,
\end{eqnarray*}
and it has a more complex, nonlinear dependence on $B$ than previously
considered. The derived magnetization and susceptibility still satisfy the same
complementarity as can be seen from Fig. \ref{fig2}. Contrary to the Ising
model, we now obtain two different domains of behavior. Above the critical
external field we have no interference and this corresponds to the disordered
phase (the visibility is low). At the point of critical field value, the
interference appears and we begin to have an ordered phase (here both
predictability and visibility contribute significantly as can be seen in
Fig.\ref{fig2}). We stress that this quantum phase transition happens at zero
temperature, and so the interference is not simply due to the equality of
Boltzmann factors introduced by the high temperatures irrespectively of the
energy eigenvalues. The interference at high temperatures does also lead to an
increase in visibility, but this has no relation to the phase transition. So,
the disorder--order transition represents a change from an effective single
slit diffraction to a double slit interference scenario. The relationship
between degeneracy and criticality is also important for classical phase
transitions as is reviewed in detail in \cite{Lieb}, and our analysis can
equally well be applied here.
\smallskip

\noindent \textbf{Conclusions.} By viewing the behavior of complex
thermodynamical systems as a double slit interference pattern we can
gain various insights into the complicated interplay between
macroscopic properties of solids. This analogy allows us to trace
this interplay directly back to the exponential factors in the
partition function which plays the same role as the amplitude
transmission function does in optics (or the density matrix in
quantum mechanics). A natural next step would be to explore more
dimensional systems and make an analogy with a multi slit
diffraction grating. It would be also beneficial to exploit this
analogy in the other direction, such as defining the free energy for
the optical case.

We believe that our work helps us develop an intuition as to which methods of
diagonalization can be successfully applied to which models and why some
methods fail for some scenarios. For example, in $1931$ Bethe used a
complicated procedure (Bethe's ansatz \cite{Bethe}) to diagonalize the one
dimensional Ising model. He concluded his work by saying that the application
of the same procedure to the two dimensional model was forthcoming in the next
paper. However, this paper never happened. From our analogy we understand now
that the one dimensional Ising model exhibits no interference phenomena and is
therefore intrinsically simple. The critical behavior existing in the two
dimensional model, on the other hand, depends crucially on the interference
which cannot be handled through Bethe's ansatz (it requires a more complicated
method invented by Onsager and reviewed in \cite{Kat}).\\
%\smallskip
\noindent{\bf Acknowledgements}. B.C.H. acknowledges EURIDICE
HPRN-CT-2002-00311.


\begin{references}
%%
\bibitem{Feynman} R. P. Feynman, {\em Statistical Physics} (Addison-Wesley, New York, 1972).
%
\bibitem{Englert} B.-G. Englert, Phys. Rev. Lett. {\bf 77}, 2154 (1996);
D. Greenberger and A. Yasin, Phys. Lett. A {\bf 128}, 391 (1988). %%
\bibitem{Beatrix} A. Bramon, G. Garbarino and B.C. Hiesmayr, Phys. Rev. {\bf 69}, 022112 (2004).
%%
\bibitem{sach99} S. Sachdev, {\em Quantum Phase Transitions}, Cambridge
Univ. Press (1999).
%%
\bibitem{Kat} S. Katsura, Phys. Rev. {\bf 127}, 1508 (1962).
%%
\bibitem{Lieb} T.D. Schulz, D.C. Mattis and E.H. Lieb, Rev. Mod.
Phys. {\bf 36}, 856 (1964). %%
\bibitem{Bethe} H. Bethe, Z. Phys. {\bf 71}, 205 (1931).

\end{references}
\end{document}